\documentclass[12pt]{article}

\def\be{\begin{equation}}
\def\ee{\end{equation}}
\def\ben{\begin{displaymath}}
\def\een{\end{displaymath}}
\def\ba{\begin{array}{c}}
\def\ea{\end{array}}
\def\p{\partial}

\newcommand{\bea}{\begin{eqnarray}}
\newcommand{\eea}{\end{eqnarray}}

\newcommand{\kt}{\rangle}
\newcommand{\br}{\langle}
\newcommand{\ed}{\end{document}}
\newcommand{\bbr}{\br\!\br}

\begin{document}

\begin{center}

{\Large \bf

Time-dependent version\\ of cryptohermitian quantum theory
  }

\vspace{9mm}

{Miloslav Znojil}

\vspace{9mm}

Nuclear Physics Institute ASCR, 250 68 \v{R}e\v{z}, Czech Republic

{znojil@ujf.cas.cz}

{http://gemma.ujf.cas.cz/\~{}znojil/}

\end{center}

\section*{Abstract}

For many quantum models an apparent non-Hermiticity of observables
just corresponds to their hidden Hermiticity in another, physical
Hilbert space. For these models we show that the existence of
observables which are manifestly time-dependent may require the use
of a manifestly time-dependent representation of the physical
Hilbert space of states.

\newpage

\section{Introduction \label{one} }

In standard textbooks one finds numerous examples of an elementary
quantum Hamiltonian $H = p^2+V(x)$ which describes a particle (or
quasi-particle) moving in a time-independent external field $V(x)$
in one dimension, $x \in I\!\!R$. The time-evolution of its wave
function $\Phi(x,t)$ is determined by the time-dependent
Schr\"{o}dinger equation
 \be
 {\rm i}\p_t\Phi(x,t) = H\,\Phi(x,t)\,.
 \label{SEt}
 \ee
Tacitly, it is understood that all of the operators of observables
${\Lambda}_j$ with $j=1,2,\ldots$ (including also the Hamiltonian
itself at $j=0$, $H ={\Lambda}_0$) are self-adjoint, $
\Lambda_j=\Lambda_j^\dagger\,$, $j = 0, 1, \ldots $. Even if the
external field becomes manifestly time-dependent, $V=V(x,t)$, the
generalized model with $H=H(t)=H^\dagger(t)$ (and, optionally, with
an appropriate experimental background for the time-dependence of
$\Lambda_1=\Lambda_1(t)=\Lambda_1^\dagger(t)$ etc) does not
necessitate any modification of the time-evolution law (\ref{SEt}).

Bender and Boettcher \cite{BB} conjectured (and, subsequently,
Dorey, Duncan and Tateo proved \cite{DDT}) that certain manifestly
non-Hermitian Hamiltonians $H= p^2+V(x) \neq H^\dagger$ may also
generate a purely real, i.e., in principle, observable spectrum of
bound-state energies. This re-attracted attention to several older
works where the similar ideas appeared in the context of field
theory \cite{Nakanishi} or relativistic quantum mechanics \cite{FV}
or nuclear physics \cite{Geyer}. Empirically, the reality of spectra
of similar Hamiltonians has been attributed to their ${\cal
PT}-$symmetry~\cite{BG}, ${\cal CPT}-$symmetry \cite{BBJ},
quasi-Hermiticity \cite{Geyer,Dieudonne,Ali} or cryptohermiticity
\cite{Smilga}. The related innovation of methods led to a new round
of perceivable progress in relativistic quantum mechanics
\cite{Alirev}, quantum cosmology \cite{wdw}, statistics \cite{Vit}
and scattering theory \cite{Jones} and even in classical
electrodynamics \cite{loran} and magnetohydrodynamics \cite{uwedva}.
In this context, the title ``Making sense of non-Hermitian
Hamiltonians" of the recent review paper \cite{Carl} written by Carl
Bender gives the name to one of the most remarkable recent projects
in theoretical physics.

In refs.~\cite{Fring,timedep} the idea has tentatively been extended
to the time-dependent non-Hermitian Hamiltonians
 \be
 H=H(t)\neq H^\dagger(t)\,.
 \label{nasi}
 \ee
Unexpectedly,  a number of conceptual difficulties has been
encountered. Serious problems have arisen, first of all, in
connection with the probabilistic and unitary-evolution
interpretation of the generalized models. For this reason, just a
very special linear time-dependence of $H(t)$  has been admitted in
\cite{Fring} and the so called quasi-stationary generalization of
this constraint has been accepted in \cite{timedep}. The resulting
theories of time-dependence with constraints looked incomplete and
deeply unsatisfactory.

In our present note we intend to reanalyze the problem. In essence,
we shall reveal that all of the specific and constrained,
quasi-stationary models are based on the same, purely intuitive and
unfounded {\em assumption} that even for all of the models with
property (\ref{nasi}) the time-dependent Schr\"{o}dinger equation
{must} remain valid in its naive, non-covariant form (\ref{SEt}). We
shall show here that after this assumption is relaxed, the theory
becomes transparent again. We shall point out that the
time-dependent non-Hermitian Hamiltonians (\ref{nasi}) may simply
{\em cease} to play the role of the generators of time evolution in
general.

In the preliminary part of our text, section \ref{IV} will summarize
a few basic mathematical features of cryptohermitian operators,
i.e., of the Hamiltonians $H$ and/or other operators of observables
$\Lambda_1, \Lambda_2,\ldots$ with real spectra which are manifestly
non-Hermitian in an {\em auxiliary} Hilbert space ${\cal H}^{(A)}$.
We shall emphasize that the reality of the respective spectra is to
be understood as a direct consequence of the standard Hermiticity
requirements imposed in {\em another, physical} Hilbert space of
states ${\cal H}^{(P)}$.

Our main results will be presented in section \ref{jedinec} where
we shall show how the cryptohermitian time-evolution law has to be
modified in order to preserve the consistency of the theory in
${\cal H}^{(P)}$. These observations will be complemented by their
brief discussion and summary in section \ref{V}.

\section{Cryptohermitian Hamiltonians 
\label{IV} }

For models with property (\ref{nasi}) a striking contrast emerges
between their innovative mathematics and conservative physics. In
\cite{Carl} Carl Bender identifies a deeper source of interest in
Hamiltonians $H \neq H^\dagger$ in the theoretical weakness of the
current practice where among all of the eligible representations of
the quantum Hilbert space of states ${\cal H}$ people most often
choose the most friendly one, viz., the space ${\cal
H}^{(A)}={I\!\!L}^2({I\!\!R})$ composed of the square-integrable
complex functions of the single real variable $x \in I\!\!R$. In
this representation the inner product between two equal-time wave
functions $\psi_a(x,t)$ and $\psi_b(x,t)$ is trivial,
 \be
  ( \psi_a,\psi_b)^{(A)} =\int_{I\!\!R}\,\psi_a^*(x,t)\,\psi_b(x,t)\,dx
  \,
  \label{usualll}
 \ee
but its use seems to exclude,  as unphysical, any complex potential
$V(x)$. Still, for many concrete non-Hermitian Hamiltonians $H =
p^2+V(x)\neq H^\dagger$ exhibiting the standard kinetic energy +
potential energy structure {\em and} leading to the real spectrum of
bound states one {\em can} admit a  complex $V(x)$. Naturally, the
representation of the physical Hilbert space of states must be
changed. Such a change of space from auxiliary to physical,
 \ben
 {\cal H}^{(A)}\ \longrightarrow\ {\cal H}^{(P)}
 \een
is usually nontrivial. At the same time, the innovated Hilbert space
${\cal H}^{(P)}$ need not necessarily be {substantially} different
from its unphysical partner ${\cal H}^{(A)}$. In the majority of
applications the latter two spaces even coincide as vector spaces
while the second one merely requires a modified definition of the
inner product using the following double integral,
 \be
   ( \psi_a,\psi_b)^{(P)} = \int_{I\!\!R^2}\,\psi_a^*(x,t)\,
 \Theta(x,x',t)\,\psi_b(x',t)\,dx\,dx'\,.
    \label{newhe}
 \ee
The time $t$ is considered fixed and the integral kernel
$\Theta(x,x',t)$ itself is usually called ``metric".

One of the most visible features shared by virtually all of the
quantum models defined within any anomalous, non-Dirac \cite{Carl}
Hilbert space ${\cal H}^{(P)}$ with metric $\Theta \neq I$ is that
the underlying Hamiltonian $H$  {\em looks} non-Hermitian,
 \be
 H \neq H^\dagger\, \ \ \ \ \ \ \ {\rm in }
 \ \ \ \ \ \
 {\cal H}^{(A)} = {\cal H}^{(unphysical)}\,.
 \label{standa}
 \ee
The same Hamiltonian $H$ remains safely self-adjoint (and, hence,
standard and physical) in ${\cal H}^{(P)}$. Unfortunately, the
definition of the Hermitian conjugation $H \to H^\ddagger$ derived
from eq. (\ref{newhe}) is more complicated and depends on the
metric,
 \be
 H = H^\ddagger=\Theta^{-1}\,H^\dagger\,\Theta^{}\,
 \ \ \ \ \ \ \ {\rm in }
 \ \ \ \ \ \
 {\cal H}^{(P)} = {\cal H}^{(physical)}\,.
 \label{standa}
 \ee
A certain complementarity is encountered between physics which is
correct in ${\cal H}^{(P)}$ and mathematics which usually proves
much easier in ${\cal H}^{(A)}$ \cite{Geyer}. Thus, one is only
allowed to speak about a loss of Hermiticity of the Hamiltonian in
the irrelevant, auxiliary space ${\cal H}^{(A)}$ with, typically,
 \ben
 \Theta^{(Dirac)}(x,x',t) = \delta(x-x')
 \een
in eq.~(\ref{newhe}). All the postulates of quantum theory remain
unchanged in the correct space ${\cal H}^{(P)}$.

\section{Time-dependent Schr\"{o}dinger equation
\label{jedinec} }

\subsection{A hermitization of the Hamiltonian \label{jedigtvrnecka} }

Inside the physical Hilbert space of states ${\cal H}^{(P)}$ let
us contemplate an {\em arbitrary} auxiliary invertible operator
$\Omega=\Omega(t)$ and replace the ``upper-case" Hamiltonian $H$
by its ``lower-case" isospectral partner
 \be
 h= \Omega\,H\,\Omega^{-1} \,. \label{mappinga}
  \ee
Whenever needed, we must apply the same mapping also to all of the
other operators of observables $\Lambda_j=\Lambda_j(t)$ in ${\cal
H}^{(P)}$. It is well known that a simple change of the basis is
obtained when $\Omega$ is chosen unitary. In an opposite direction
we now intend to choose such a non-unitary map $\Omega$ that the
resulting new representation $h=h(t)$ of our Hamiltonian becomes
Hermitian,
 \be
 h=h(t)=\Omega(t)\,H(t)\,\Omega^{-1}(t)  = h^\dagger(t)\,
 =\left [\Omega^{-1}(t)\right ]^\dagger\,H(t)\,\Omega^\dagger(t)
 \,.
 \label{terestanda}
 \ee
This must be complemented by an observation that {\em both} the
representations $H(t)$ and $h(t)$ of the energy-operator {\em need
not} play the role of the generator of the time evolution {\em
simultaneously}. There exists no mathematical or physical
principle which would force us to insist on the validity of
eq.~(\ref{SEt}) when $H \neq H^\dagger$. We are allowed to
restrict our attention to the lower-case generators $h(t)$ of the
safely unitary time-evolution and to the related lower-case wave
functions defined by the integral containing the kernel of
$\Omega$,
 \be
 {\varphi}(x,t) =
   \int_{I\!\!R}\,
 \Omega(x,x',t)\,\Phi(x',t)\,dx\,dx'\,.
 \label{SEtauvej}
 \ee
When we use the current Dirac bra-ket notation the latter relation
can be abbreviated as $ |\varphi(t)\kt=\Omega^{}(t)\,|\Phi(t)\kt$.
Thus, we may characterize the state of our physical quantum
system, at any time $t$, by the very standard elementary projector
 \be
 \pi(t) =
 |\varphi(t)\kt\,\frac{1}{\br \varphi(t)|\varphi(t)\kt}\,
 \br \varphi(t)|
 \,.
    \label{lonen}
 \ee
The evolution of this expression in time is controlled by the usual
time-dependent Schr\"{o}dinger equation
 \be
 {\rm i}\p_t|\varphi(t)\kt = h(t)\,|\varphi(t)\kt\,.
 \label{SEtau}
 \ee
For any state $
  \varphi(x,t)=
 \br x |\varphi(t)\kt
 $
prepared at $t=0$ and measured at  $t>0$ the operator $h(t)$ plays
the role of its self-adjoint generator of evolution in time.

Our knowledge of this generator enables us to introduce the
evolution operator $u(t)$ defined by the following operator
alternative to eq.~(\ref{SEtau}),
 \be
  {\rm i}\partial_t u(t)=h(t)\,u(t)\,.
 \label{seh}
 \ee
The formal solution of eq.~(\ref{SEtau}) then reads
 \be
 |\varphi(t)\kt \ \ = \ u(t)\,|\varphi(0)\kt\,
 \label{timeq}
 \ee
and, obviously, it satisfies the identity
 \ben
 \br \varphi(t) \,|\,
 \varphi(t)\kt=
 \br \varphi(0) \,|\,
 \varphi(0)\kt\,.
 \een
This identity is a guarantee that the evolution of the system is
unitary.

\subsection{The doublet of pull-backs of wave function
 \label{jedivrnecka} }

For a sufficiently general kernel $\Omega(t)$ in
eq.~(\ref{SEtauvej}) the Hermitian representation $h(t)$ of the
Hamiltonian is a complicated integro-differential operator. A
return to  $H(t)=p^2+V(x,t)\neq H^\dagger(t)$ makes good sense,
therefore. One of the most immediate consequences of the resulting
parallel work in ${\cal H}^{(A)}$ and ${\cal H}^{(P)}$ is that  we
must carefully distinguish between the respective bra-vectors. In
the usual Dirac notation, for example, Mostafazadeh \cite{Ali}
recomends the two-letter notation where $
   \left [ \
 |\Phi(t)\,\kt\
   \right ]^\ddagger
   \,\equiv\,\br \Psi(t)|$
and where, in the light of eq.~(\ref{newhe}),
 \ben
   \br x|\Phi_b(t)\,\kt  \,\equiv\,
   \psi_b(x,t) \in {\cal H}^{(P)}\,,
   \ \ \ \ \ \ \ \ \ \ \ \ \ \ \ \ \ \ \ \ \ \ \ \
   \een
   \ben
   \br\Psi_a(t)|x \kt
   \,\equiv\,
   \int_{I\!\!R}\,\psi_a^*(y,t)\,
 \Theta(y,x,t)\,dy
 \in \left [{\cal H}^{(P)}\right]'
 \,.
    \label{newherg}
 \een
This convention well reflects the fact that in ${\cal H}^{(P)}$
there emerge two different pull-backs of the single wave function
(\ref{timeq}), viz.,
 \be
 |\Phi(t)\kt=\Omega^{-1}(t)\, |\varphi(t)\kt\,,
 \ \ \
 \br \Psi(t)\,|=\br\varphi(t)\,|\,\Omega(t)\,.
 \label{lykra}
 \ee
One should emphasize that in spite of being marked by the two
different Greek letters, these symbols still represent the {\em
same} state of our physical system in question. Formally, this
description is provided by the upper-case pull-back of the projector
$\pi(t)$ given by eq.~(\ref{lonen}),
 \be
 \Pi(t) =
 |\Phi(t)\kt\,\frac{1}{\br \Psi(t)|\Phi(t)\kt}\,
 \br \Psi(t)|
 \,.
 \ee
%
%
%
The new projector $\Pi(t)=\Pi^\ddagger(t)$ remains non-Hermitian in
the unphysical space ${\cal H}^{(A)}$ and its construction requires
the knowledge of the {\em pair} of time-dependent functions or
vectors (\ref{lykra}). As long as $\Omega^{-1}(t)\neq
\Omega^\dagger(t)$, these two complex functions of $x$ are different
in general.

Our knowledge of the time-dependence of the latter two functions is
a remarkable consequence of the construction. Schr\"{o}dinger
eq.~(\ref{SEtau}) and some elementary algebra lead to the
right-action evolution rule
 \be
 |\Phi(t)\kt=U_R(t)\, |\Phi(0)\kt\,,\ \ \ \ \ \
 U_R(t)=\Omega^{-1}(t)\,u(t)\,\Omega(0)
 \ee
accompanied by its left-action parallel
 \be
 |\Psi(t)\kt=U_L^\dagger(t)\, |\Psi(0)\kt\,,\ \ \ \ \ \
 U_L^\dagger(t)=\Omega^\dagger(t)\,u(t)\,
 \left [\Omega^{-1}(0)\right ]^\dagger\,.
 \ee
The respective non-Hermitian analogues of the Hermitian
evolution-operator rule~(\ref{seh}) are obtained by the elementary
differentiation and insertions yielding
 \be
 {\rm i}\partial_t U_R(t)=
 -\Omega^{-1}(t)
 \left [{\rm i}\partial_t\Omega(t)
 \right ]\, U_R(t)+H(t)\, U_R(t)\,
 \ee
and
 \be
 {\rm i}\partial_t U_L^\dagger(t)=
 H^\dagger(t)\, U_L^\dagger(t)
 +
 \left [{\rm i}\partial_t\Omega^\dagger(t)
 \right ]\,
 \left [
 \Omega^{-1}(t)
 \right ]^\dagger\, U_L^\dagger(t)\,.
 \ee
We achieved our goal. In the language of mathematics the latter
doublet of operator equations offers a differential-equation
simplification of the equivalent integro-differential lower-case
Schr\"{o}dinger eq.~(\ref{seh}). Thus, the role of the complicated
lower-case representation $u(t)$ of the evolution operator is
transferred to its two upper-case maps which offer a consistent
description of quantum dynamics in ${\cal H}^{(P)}$.

\section{Discussion \label{V} }
%
%

Several misunderstandings concerning the pull-backs of wave
functions have recently been encountered in a series of comments and
replies on arXive \cite{comments}. There, a few unexpected
properties of the generalized quantum time-evolution equations have
been discussed, with the final clarification of the puzzle presented
in the preliminary preprint version \cite{which} of our present
paper. It makes sense, therefore, to perform an independent check of
what happens with the norm $ \br \Psi(t)\,|\,\Phi(t)\kt $ of a given
state which evolves with time in ${\cal H}^{(P)}$. The elementary
differentiation confirms that
 \bea
 {\rm i}\partial_t\br \Psi(t)\,|\,\Phi(t)\kt
 ={\rm i}\partial_t
 \br \Psi(0)\,|\,U_L(t)\,U_R(t)  \,|\,\Phi(0)\kt=
 \nonumber
 \\
 =\br \Psi(0)\,|\,
 \left [ {\rm i}\partial_t
 U_L(t)
 \right ]
 \,U_R(t)  \,|\,\Phi(0)\kt
 +\br \Psi(0)\,|\,U_L(t)\,
 \left [ {\rm i}\partial_t
 U_R(t)
 \right ]
   \,|\,\Phi(0)\kt=
 \nonumber
   \\
 \nonumber
 = \br \Psi(0)\,|\, U_L(t)\,
 \left [ -H(t)+\Omega^{-1}(t)
 \left [{\rm i}\partial_t\Omega(t)
 \right ]
 \right ]
 \,U_R(t)  \,|\,\Phi(0)\kt+\\
 +\br \Psi(0)\,|\,U_L(t)\,
 \left [ H(t)
  -\Omega^{-1}(t)
 \left [{\rm i}\partial_t\Omega(t)
 \right ]
 \right ]\,U_R(t)
   \,|\,\Phi(0)\kt=0\,.
 \nonumber
   \eea
We see that irrespectively of the mapping $\Omega$ the norm does not
vary so that the time-evolution of the system is unitary also  by
this check. It reconfirms that the naive picture of the
time-evolution as generated by the non-Hermitian Hamiltonian $H(t)$
is incomplete.

In our present brief paper we were more constructive in showing that
whenever $H \neq H^\dagger$, the time evolution must in general be
prescribed by a {\em pair} of modified Schr\"{o}dinger equations.
With the purpose of making this argument fully explicit, let us
abbreviate $\partial_t\Omega(t)\equiv \dot{\Omega}(t)$ and write
down the following explicit specification of the time-evolution
generator in ${\cal H}^{(P)}$,
 \be
 H_{(gen)}(t)=H(t) -{\rm i}\Omega^{-1}(t)
 \dot{\Omega}(t)\,.
 \label{homer}
 \ee
It is remarkable that this operator enters {\em both} the updates of
the Schr\"{o}dinger equation for wave functions in ${\cal H}^{(P)}$,
 \bea
 {\rm i}\partial_t|\Phi(t)\kt
 =H_{(gen)}(t)\,|\Phi(t)\kt\,,
 \label{SEA}\\
 {\rm i}\partial_t|\Psi(t)\kt
 =H_{(gen)}^\dagger(t)\,|\Psi(t)\kt\,.
 \label{SEbe}
 \eea
Such a confirmation of the overall unitarity of the evolution comes
at a very reasonable cost of the covariant redefinition $H \to
H_{(gen)}$ of its generator.

We can summarize that the adequate and fairly universal picture of
quantum  dynamics can be reinstalled in its upper-case
cryptohermitian (i.e., typically, less non-local and technically
simpler) representation provided only that one admits that the
time evolution is {\em not necessarily} generated by the naive,
non-covariant map $H(t)$ of a physical self-adjoint Hamiltonian
$h(t)$. This confirms that the ``traditional" Schr\"{o}dinger eq.
(\ref{SEt}) may cease to be valid in general. The existence of
papers like \cite{timedep,Scolar} as well as  of several
unpublished comments on the web \cite{comments} indicates that
this observation is nontrivial and that it can perceivably extend
the range of applicability of cryptohermitian models in quantum
theory.

\subsection*{Acknowledgement}

Work supported by GA\v{C}R, grant Nr. 202/07/1307, Institutional
Research Plan AV0Z10480505 and by the M\v{S}MT ``Doppler
Institute" project Nr. LC06002.

\newpage

\subsection*{Appendix: Quasi-stationarity constraint  }

One of the unexpected mathematical benefits of the form of operator
(\ref{homer}) is that it is the same for both its left and right
action. Still, its decisive appeal lies in the universality of its
description of physics  where the cryptohermitian observables and,
in particular, cryptohermitian Hamiltonian operators $H=H^{(CH)}$
are allowed to be arbitrary (or at least arbitrary analytic)
functions of time $t$,
 \be
 H^{(CH)}(t)= H_{(0)}+t\,H_{(1)}+t^2\,H_{(2)}+\ldots\,.
 \label{CH}
 \ee
Several aspects of the underlying idea of having the manifestly
time-dependent metric $\Theta = \Theta(t)$ (i.e., the time-dependent
representation of our Hilbert space of states) may look slightly
counterintuitive \cite{comments}. For this reason, several authors
\cite{timedep,Scolar} tried to restrict the class of the
cryptohermitian time-dependent models by the so called
quasi-stationarity constraint
 \be
 \Theta =\Theta^{(QS)} \neq \Theta(t)\,.
 \label{QS}
 \ee
Such a postulate is in fact rather unfortunate. In an attempt of
leaving the form of eq.~(\ref{SEt}) unchanged, it practically
eliminates the possibility of a consistent application of quantum
mechanics to the majority of systems with a sufficiently nontrivial
time-dependence of its observables $\Lambda_j(t)$.

In the light of empirical results of ref.~\cite{Fring}, the latter
statement can even be strengthened and made more quantitative
since in the generic quasi-stationary case the infinite Taylor
series of eq.~(\ref{CH}) {\em must} degenerate to the linear
polynomial
 \be
 H^{(QS)}(t)= H_{(0)}+t\,H_{(1)}\,.
 \label{LP}
 \ee
The preliminary, two-by-two matrix illustration of such a key
drawback resulting from assumption (\ref{QS}) can be found in
ref.~\cite{timedep}. Here, we just intend to complement this
example by a less model-dependent demonstration that the linearity
constraint (\ref{LP}) is generic, for the finite-dimensional
models at least.

In the first step of our proof we accept the assumption that a given
$N$ by $N$ cryptohermitian Hamiltonian $H^{(CH)}$ (with $N \leq
\infty$) is quasi-stationary, time-dependent and cryptohermitian,
i.e.,
 \be
 \left [H^{(CH)}(t)
 \right ]^\dagger=\Theta^{(QS)}\,H^{(CH)}(t)\,\left [
 \Theta^{(QS)}\right ]^{-1}\,.
 \label{ohonn}
 \ee
With a constant, time-independent  metric $\Theta^{(QS)}$ this
requirement can be rewritten as an {\em infinite} family of
equations to be satisfied by the coefficients in eq.~(\ref{CH}),
 \be
 H_{(m)}^\dagger\,\Theta^{(QS)} =\Theta^{(QS)}\,H_{(m)}\,
 \ \ \ \ \ \ m = 0, 1, \ldots\,.
 \label{hoohonn}
 \ee
Up to exceptional cases which will not be discussed here, all of the
individual $N$ by $N$ matrix coefficients $H_{(m)}\neq
H_{(m)}^\dagger$ in eq.~(\ref{CH}) may be assumed diagonalizable,
 \ben
 H^{(CH)}_m
 = \sum_{j=1}^{N}\,|\Phi_{m,j}\,\kt\,\varepsilon_{m,j}\,
 \br \Psi_{m,j}\,|\,.
 \een
Each choice of $m=0,1,\ldots$ specifies, in general, a different
biorthonormalized set of vectors together with a different real
$N-$plet of eigenvalues $\varepsilon_{m,j}$ with $j=1,2,\ldots,N$.

At this moment we remind the readers that at any subscript
$m=0,1,\ldots$ we may specify and construct the $N-$plet of the
``left eigenvectors" $|\Psi_{n,j}\kt$ as a set of  biothonormalized
eigenvectors of the conjugate matrix $H_{(m)}^\dagger$. In terms of
these vectors \cite{foot} we may write the Mostafazadeh's \cite{Ali}
most general spectral expansion
 \be
  \Theta^{(QS)}=\sum_{n=1}^N\,|\,\Psi_{0,n}\kt\,\kappa_{0,n}
  \br \Psi_{0,n}|
 \,.
 \label{srf0}
 \ee
At any finite $N$ this formula describes all the metrics
compatible with eq.~(\ref{hoohonn}) at $m=0$. They depend on  $N$
free parameters $\kappa_{0,n}$ which must be real and positive
\cite{SIGMA}.

In the next step of our proof we contemplate the overlap matrix
 \ben
 {\cal A}_{jk}=\br \Psi_{0,j}\,|\,\Phi_{1,k}\kt\,,
 \een
and deduce that
 \ben
 {\cal B}_{jk}=\br \Psi_{1,j}\,|\,\Phi_{0,k}\kt
 =\left ({\cal A}^{-1}\right )_{jk}\,.
 \een
Then, the insertion of eq.~(\ref{srf0}) and the use of the two
diagonal real matrices $T$ (with elements $T_{jj}={\kappa_{0,j}}$)
and $F$ (with elements $F_{jj}=\varepsilon_{1,j}$) transforms
eq.~(\ref{hoohonn}) into a remarkably compact matrix relation at the
next subscript $m=1$,
 \be
  T\,{\cal A}\,F\,{\cal A}^{-1}=
 \left ({\cal A}^{-1}
 \right )^\dagger\,F\,{\cal A}^\dagger\,T
 \,.
 \label{srf11}
 \ee
The first line of this relation has the form of a vectorial identity
 \ben
 \left (
 M_{11}\kappa_{0,1}\,,
 M_{12}\kappa_{0,1}\,,\ldots\,,M_{1N}\kappa_{0,1}
 \right )=
 \left (
 M_{11}^*\kappa_{0,1}\,,
 M_{21}^*\kappa_{0,2}\,,\ldots\,,M_{N1}^*\kappa_{0,N}
 \right )\,
 \een
where all the matrix elements $M_{jk}$ are known. In the generic
case and up to an irrelevant overall factor this relation defines
all the parameters $\kappa_{0,n}$ in the metric (\ref{srf0}),
therefore. The rest of eq.~(\ref{srf11}) is redundant. This
observation may be read either as a proof of the non-existence of
$\Theta^{(QS)}$ for a general ``dynamical input" $H_{(1)}$ or,
alternatively, as a set of nontrivial compatibility conditions which
must be imposed upon the ``acceptable" matrices $H_{(1)}$ in
eq.~(\ref{LP}).

We see that even the linear time-dependence of the Hamiltonian
characterized by the matrix coefficient $H_{(1)}$ is not
arbitrary. Moreover, once we choose its most general form we are
left with no free parameters which could guarantee the
compatibility between our quasi-stationary metric $\Theta^{(QS)}$
and any higher-order coefficient $H_{(m)}$ at some $m \geq 2$ in
Taylor series (\ref{CH}). We may summarize that in the
cryptohermitian quantum models restricted to the quasi-stationary
regime the quadratic and higher-power time-dependence of its
observables can only occur as a very exceptional, fine-tuned
phenomenon.


\begin{thebibliography}{99}
\bibitem{foot}
Marginally let us note that in \cite{SIGMA} we recommended an
alternative and, perhaps, slightly more compact and transparent
notation convention where the standard Dirac abbreviation
$|\Phi_{m,j}\,\kt\,\equiv\,|{m,j}\,\kt$  is complemented by the
``double-bra" marking for $\left (|\Phi_{m,j}\,\kt\right )
^\ddagger=\br \Psi_{m,j}\,|\,\equiv\,\bbr {m,j}\,|$.
\end{thebibliography}
\end{document}